# Capillary Condensation in Nanogaps: Nucleation or Film Coalescence?

Gentrit Zenuni,[1] Ari Laaksonen,[2,3] Robin H. A. Ras,[1,4] and Ali Afzalifar[1,4,*]

[1]*Department of Applied Physics, Aalto University, P.O. Box 15600, 00076 Aalto, Espoo, Finland*
[2]*Finnish Meteorological Institute, 00560, Helsinki, Finland*
[3]*Department of Technical Physics, University of Eastern Finland, 70211 Kuopio, Finland*
[4]*Center of Excellence in Life-Inspired Hybrid Materials (LIBER), Aalto University, Espoo, Finland*

Nucleation and film coalescence represent two fundamentally different pathways for capillary condensation. Yet, both have so far been proposed as the processes driving the condensation in nanometric confinements, leading to a long-standing and overlooked ambiguity. Here, we delineate the dichotomy between these mechanisms and test their validity using an experimental method capable of absolute distance measurement during capillary condensation. We show that the molecular content of the capillary meniscus given by the first nucleation theorem is far smaller than what the confinement geometry and the Kelvin equation require. In contrast, the analysis based on film coalescence reproduces the experimental observations and describes the final meniscus formation as a barrierless process, while allowing for an intermediate, first-order-like film-thickening transition prior to the meniscus formation.

*Introduction*—The general understanding about capillary condensation in nanometric confinements is that, for instance in the case of water, depending on the ambient humidity and the surface wettability, when the separation between two surfaces becomes smaller than a critical distance, a capillary bridge or meniscus forms between them. Most studies attribute the formation of the capillary bridge to nucleation [1–7], so based on the nucleation theory [8], the process must be stochastic and requires surmounting a free energy barrier, i.e. nucleation work [Fig. 1(a)]. The nucleation work passes through a maximum corresponding to the critical bridge size with molecular content $n^*$. Only bridges larger than $n^*$ are stable and grow larger, while smaller ones decay and evaporate. In contrast, a few works have explained the capillary condensation in the framework of instability and coalescence of adsorbed liquid films [9–12]. This view entails prior presence of liquid layers on the surfaces, and thus explains capillary bridge formation in a way which can be barrierless and deterministic [Fig. 1(a)].

Although both nucleation and, to a lesser extent, coalescence have been examined separately in earlier works, their distinction as two fundamentally different pathways for capillary condensation has not been investigated. Here, we directly address the question of whether capillary condensation in slit pores proceeds by nucleation or by film coalescence, delineate these mechanisms, and experimentally test their validity, thereby clarifying a long-standing and overlooked ambiguity in the field.

*Experimental method*—We aimed to achieve absolute, rather than averaged, measurement of the critical distance during capillary condensation. Atomic force microscopy (AFM) might appear to afford a convenient method for this purpose, as it has been commonly used for studying capillary condensation [5–7,13,14]. However, the inherent tip oscillation in AFM renders the measured distance an averaged quantity. Furthermore, the oscillation amplitude is amplified at the tip's end [15] introducing significant uncertainty when the absolute tip-substrate distance is needed. To overcome this limitation, we developed a method similar to scanning tunneling microscopy (STM), but distinct in its voltage–current range and the nature of the measured currents. A few studies have utilized capillary bridge formation between an STM tip and a wet sample for imaging biological specimens [16–18]. However, the capillary bridge itself has not been the focus of a dedicated STM-based investigation, mainly because the presence of water films is generally regarded a nuisance in non-vacuum STM studies rather than a subject of interest.

Our method involves a temperature- and humidity-controlled chamber where a tip is brought close, with steps of 0.4 Å, to a substrate of highly-doped *n*-type Si(100) to form a capillary bridge in the tip-substrate gap. The substrate has a resistivity of 1-5 mΩ.cm and is covered by a ~2 nm film of native $SiO_2$. Just before the measurements, the substrate was rendered almost perfectly wettable (with apparent water contact angle $\theta < \sim 3°$) by Ar-plasma treatment.

A 30 mV bias is applied between the tip and substrate, and the onset of the capillary bridge formation is registered when the first sudden jump in the current from the noise level (~0.35 fA = $2.2 \times 10^3$ e⁻/s) to tens of fA occurs [Fig. 1(b)]. This current is insensitive to the tip-substrate distance and disappears for $S \approx 0$, where $S$ is the saturation ratio or relative humidity, because it is attributed to ionic/faradaic conduction through the water meniscus involving low-level charge transfer at the Pt–Ir/water/$SiO_2$ interfaces. Upon further closing the gap, a substantial and highly-distance-sensitive increase in the current occurs as the tip mechanically contacts the $SiO_2$ film. This may occur via direct tunneling through locally thinned oxide regions, trap-assisted conduction, or localized oxide disruption under contact, all of which lead to a steep, distance-sensitive rise in current. The difference between these two jumps in the current gives the critical height of the tip-substrate gap $H_c$ at which the capillary bridge forms [Fig. 1(b)].

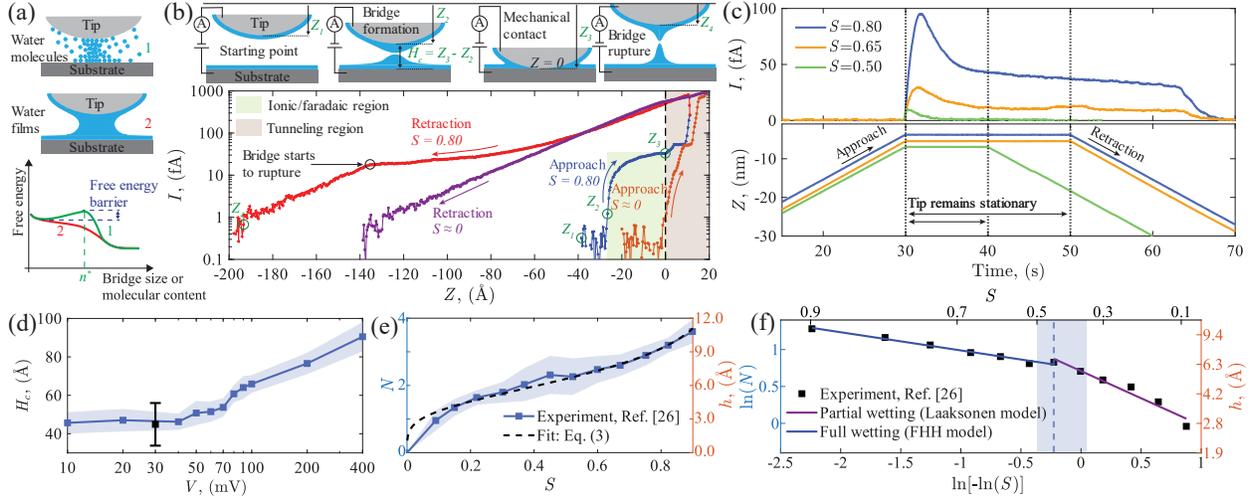

FIG 1. (a) Simplified illustration of a capillary condensation process and free energy change via (1) nucleation and (2) film coalescence. (b) Experimental method schematic (top), and representative current–distance measurements for approach-retraction cycles at $S = 0.8$ and $S \approx 0$ (bottom). (c) Current–time (top) and distance–time (bottom) measurements to examine the ionic/faradaic current. (d) Dependence of $H_c$ on tip–substrate bias for $S = 0.8$, each point represents 10 measurements (100 at 30 mV). The shaded area/error bar indicates standard deviations. (e) and (f) Experimental water adsorption isotherm on native $SiO_2$ in $N$—$S$, and $\ln(N)$—$\ln[-\ln(S)]$ space, where $N$ is the number of adsorbed monolayers. FHH-Laaksonen model fitting identifies the wetting transition near $S \approx 0.45$.

To further examine the current behavior upon bridge formation, we performed a set of special measurements at different $S$. Once bridge formation is detected (by registering a 5-fA threshold), the tip was parked to maintain a fixed tip–substrate distance for 20, 20 and 10 s at $S = 0.80, 0.65$ and $0.5$, respectively [Fig. 1(c)]. The current shows an initial transient jump, identical to the first current jump discussed before, and then stabilizes at a nearly constant value, persisting even as the tip is retracted. The current only begins to decrease when the capillary bridge ruptures. This behavior demonstrates that the water meniscus provides a stable, distance-insensitive conductive pathway; accordingly, reducing $S$ shrinks the bridge and lowers the current.

We used a blunt Pt-Ir tip ($r_{tip} \approx 400$ nm) to warrant analyzing the tip–substrate gap geometry as a symmetric slit pore (height $\ll$ width), and neglecting deformation and possible compositional changes of the tip apex caused by repeated mechanical contacts. Water adsorption on Pt-Ir amounts to only 1–3 monolayers at $S = 0.8$ [19], indicating a lower wettability than $SiO_2$. Therefore, in the experiments, the first five approach–retraction cycles were discarded to allow equilibration with surface moisture, thereby approximating the symmetry assumed in the analysis.

Our voltage sensitivity analysis shows that $H_c$ is insensitive to tip–substrate bias for $V \leq 70$ mV [Fig. 1(d)]. Moreover, at a 1 nm gap, a 30 mV bias corresponds to a field of only 0.03 V/nm—several orders smaller than the field thresholds required for water field dissociation [20], anodic oxidation [21,22], or altering water orientation and density [23–25]. Also, such a low bias keeps currents $< \sim 100$ fA, minimizing electrochemical reactions.

*A thought experiment*—To simplify our question, we propose a thought experiment with two limiting scenarios: (1) when one or both surfaces are covered by water films, and (2) when both are dry. In the first scenario, logically the water films must be involved in the meniscus formation. Then, the remaining task is to ascertain whether the meniscus forms by nucleation or by coalescence. If nucleation can be ruled out, coalescence necessarily follows. In the second scenario, the surfaces are dry, and coalescence is therefore excluded. This makes our task simpler, because only showing that bridge formation is impossible would rule out nucleation as well. When this rejection co-occurs with coalescence being the case in the first scenario, it would eliminate nucleation as a possible mechanism for capillary condensation.

To implement the conceptual framework outlined above, it was first necessary to define the range of $S$ for wet and dry surface conditions. Therefore, experimental water adsorption isotherm on native $SiO_2$ [25], Fig. 1(e), was analyzed by combining Frenkel–Halsey–Hill (FHH) [27–29] and Laaksonen [30] models, Fig. 1(f). The former describes multilayer adsorption (full wetting) at high humidities, whereas the latter accounts for droplet-like adsorption (partial wetting) at low humidities. Fitting the data in $\ln(N) - \ln(-\ln(S))$ space locates the transition near $S \approx 0.45$. Hence, we conclude that for $0.50 \leq S$ the surface is uniformly covered by a continuous film, while for $S \leq 0.35$ it only supports discontinuous water islands.

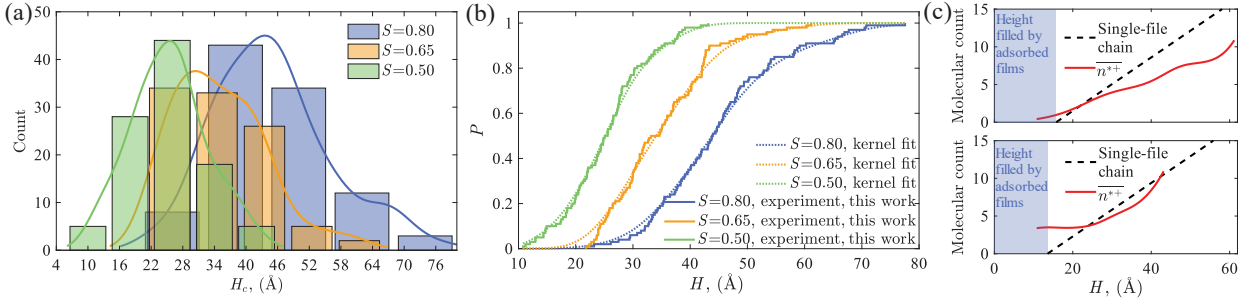

FIG 2. (a) Histograms of measured $H_c$. (b) The survival probability as a function of $H$. (c) Estimated number of molecules forming the bridge from Eq. (2), for $S = 0.80$ vs. $0.65$ (top) and $S = 0.65$ vs. $0.50$ (bottom), compared with number of molecules required for a single-file chain to bridge the remaining gap $H - 2h$ in the pore.

*The first scenario*—We measured $H_c$ at 100 locations spaced 4 μm apart on the substrate for $S = 0.50, 0.65,$ and $0.80$ [Fig. 2(a)]. For the sake of argument, we assume that capillary condensation in the tip–substrate gap is a nucleation process. Therefore, we can apply the first nucleation theorem [31], as a general, model-independent approach, to calculate the molecular content of the critical bridge $n^*$ at fixed $S$, $H$ and temperature $T$. Random nucleation events follow a Poisson distribution, thus the survival probability, the probability that no bridge has formed, decays exponentially with time $t$ as $P = \exp(-Jt)$ [8], where $J$ (s$^{-1}$) is the time-independent meniscus nucleation rate. Experimentally, $P$ is measurable as

$$P = \frac{M_H}{M_{tot}}, \quad (1)$$

where $M_{tot}$ ($= 100$) is the total number of trials, and $M_H$ is the number of trials with no bridge formed until the tip reaches $H$ [Fig. 2(b)]. Then, applying the first nucleation theorem, we define an upper-bound estimate for the molecular content of the critical bridge as

$$\overline{n^{*+}} = \frac{d \ln[-\ln(P)]}{d \ln(S)}, \quad (2)$$

in the sense that $n^*$ must be smaller than $\overline{n^{*+}}$. The derivatives in Eq. (2) were computed using the data pairs $S = 0.80$ vs. $0.65$, and $S = 0.65$ vs. $0.50$ [Fig. 2(c)]. The derivation of Eq. (2) and the justification of its applicability to our measurements are provided in Appendix A.

We know that a part of the tip-substrate height is occupied by the thickness of the adsorbed film $h$ on the substrate as given by the experiment [26], Fig. 1(e). For simplicity, we assumed an equal film thickness on the tip surface, so $2h$ of the gap height is filled with water films. Even under this assumption, $\overline{n^{*+}}$ barely amounts to enough molecules to form a single-file, one-dimensional chain across the remaining distance $H_c - 2h$. This is a physically implausible configuration for a capillary bridge, particularly given the large $r_{tip}$. Moreover, comparing $\overline{n^{*+}}$ for $S = 0.80$ vs. $0.65$ with that for $S = 0.65$ vs. $0.50$ reveals that the inferred bridge size increases as humidity decreases [Fig. 2(c)]. However, the Kelvin equation predicts smaller equilibrium bridges at lower humidity.

TABLE I. Estimated bridge dimensions and molecular content $\tilde{n}$ (assuming a catenoid geometry) from the Kelvin equation. Dimensions are in Å. Values marked † are linearly interpolated between $S = 0.8$ and $0.65$, and $S = 0.65$ and $0.5$ for comparison with $\overline{n^{*+}}$ in Fig. 2(c).

| $S$ | $\overline{H_c}$ | $h$ | $\overline{H_c} - 2h$ | $-r_1$ | $r_2$ | $\tilde{n}/10^3$ |
|---|---|---|---|---|---|---|
| 0.8 | 44.8 | 8.7 | 27.3 | 23.9 | 437.5 | 6.3 |
| 0.725 | 39.9† | 7.8 | 24.1 | 16.6 | 364.4 | 3.8 |
| 0.65 | 34.9 | 7.2 | 20.4 | 12.4 | 289.1 | 2.0 |
| 0.575 | 30.1† | 6.7 | 16.6 | 9.6 | 277.8 | 1.6 |
| 0.5 | 25.3 | 6.3 | 12.61 | 7.7 | 248.2 | 1.0 |

To compare $\overline{n^{*+}}$ with a realistic geometric estimate of the bridge, we use the Kelvin equation. Experimental evidence confirms the Kelvin equation validity for cyclohexane capillary bridges with radii as small as 4 nm [32], and in the case of homogeneous nucleation of water, it predicts values comparable with the first nucleation theorem's results, down to subnanometric clusters [33–35]. The Kelvin equation gives the mean radius of curvature of the capillary bridge $r_K$ as $\frac{1}{r_K} = \frac{1}{r_1} + \frac{1}{r_2} = \frac{kT \ln(S)}{\sigma_{lv} v}$ (see [32], and references 1-5 therein), where $r_1$ and $r_2$ are the first and second principal $\sigma_{lv}$ is the liquid-vapor surface tension and $v$ is molecular volume. When $r_2 \ll r_{tip}$ and $\cos(\theta) \approx 1$, we have $r_1 \cong$ $r_k$ and $r_2 \cong 2\sqrt{r_{tip} r_k}$ [36]. Table I shows, in contrast to $\overline{n^{*+}}$, the values of $-r_1$ are comparable to the reduced gap $\overline{H_c} - 2h$ ($\overline{H_c}$ is the mean of measured $H_c$) and show a consistent trend vs. $S$. The discrepancies between $\overline{H_c} - 2h$ and $-r_1$ are expected, since the Kelvin equation alone does not account for the effects of surface forces and interfacial transition regions [9,37]. The estimated molecular content of the bridge $\tilde{n}$ is also orders of magnitude larger than $\overline{n^{*+}}$. It is important to note that in nucleation the Kelvin equation gives the critical radius at unstable equilibrium (a free energy maximum), whereas in capillary condensation it defines a stable meniscus curvature corresponding to a global free energy minimum.

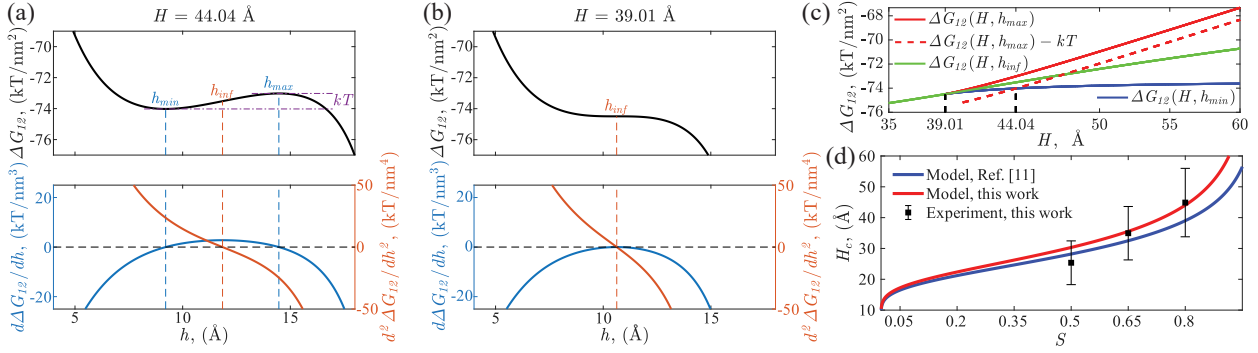

FIG 3. (a) and (b) Free energy change and its derivatives in transition from an empty pore to one containing two adsorbed films where $S = 0.8$, plotted for one film as a function of thickness $h$ at pore heights $H =$ 44.04 and 39.01 Å. (c) Extrema of $\Delta G_{12}$ as a function of $H$, showing the gradual disappearance of the free energy barrier to film thickening as $H$ decreases (note $H$ values corresponding to (a) and (b)). (d) Model predictions of $H_c$ by [11] and by this work are compared with experiment.

This analysis provides strong experimental evidence that the molecular content predicted using nucleation theorem is far smaller than what the actual meniscus geometry requires. Consequently, the meniscus is unlikely to form via nucleation. Since, in our framework, nucleation and film coalescence are mutually exclusive pathways, the results unambiguously identify coalescence as the most probable mechanism for capillary condensation.

*Explaining the first scenario based on coalescence model*—We apply the model of Churaev et al. [11] (see also [9]) which uses the film disjoining pressure to explain bridge formation in a slit pore as a result of films' instability and merging when $H \leq H_c$. The disjoining pressure of an adsorbed film on a substrate, or equivalently when $H \to \infty$, is [38]

$$\Pi(h, \infty) = \Pi_0^h \exp(-h/h_m) = -\frac{kT}{v}\ln(S), \quad (3)$$

where $\Pi_0^h$ is the characteristic strength of the hydration force field, and $h_m$ is the decay length of the field. By fitting Eq. (3) to the experimental adsorption isotherm of water on $SiO_2$; $\Pi_0^h = 1468.2$ MPa and $h_m = 2.25$ Å are obtained (R-squared = 0.98) [Fig. 1(e)]. When the pore height becomes sufficiently narrow, the mutual attraction of the opposing films contributes an additional term to the disjoining pressure as [11]

$$\Pi(h, H) = \Pi_0^h \exp(-h/h_m) + \frac{2B_1}{(H-2h)^3}, \quad (4)$$

where $B_1 = 4.8 \times 10^{-20}$ J [10] is the Hamaker constant of water film-film interactions across the unfilled portion of the pore. This term formulates the driving force for the mechanical instability and collapse of the films.

Using thermodynamics of thin-film equilibrium [39], the Gibbs free energy change per unit area for each film in the transition from (1) an empty pore to (2) a pore containing two adsorbed films reads

$$\Delta G_{21} = -\int_0^h \Pi(h', H)\, dh' - \frac{h}{v} kT \ln(S), \quad (5)$$

for derivation see Appendix B. Differentiating $\Delta G_{21}$ with respect to the film thickness yields

$$\frac{d\Delta G_{21}}{dh} = -\frac{kT}{v}\ln(S) - \Pi(h, H), \quad (6)$$

$$\frac{d^2 \Delta G_{21}}{dh^2} = -\frac{d\Pi(h, H)}{dh}. \quad (7)$$

Fig. 3(a) shows that for $H = 44.04$ Å at $S = 0.8$, $\Delta G$ exhibits a local minimum at $h_{min}$ (corresponding to a metastable equilibrium film thickness), a global maximum at $h_{max}$ (corresponding to an unstable equilibrium), and an inflection point at $h_{inf}$. As $H$ decreases, the extrema move progressively toward the inflection point, and at the critical height $H = 39.54$ Å the first and second derivatives of $\Delta G$ simultaneously vanish at a single film thickness [Fig. 3(b) and (c)]. This convergence defines a critical point at $H_c$ where the metastable region vanishes, and the system becomes critically unstable. This point is of particular importance, as it signifies the spinodal limit of film stability. Beyond this point, the restoring force against film thickness fluctuations vanishes, allowing local fluctuations to grow spontaneously until the opposing films merge and fill the pore. This continuous transition in film thickness is analogous to a second-order phase transition. In the original model [11] (see also [40]), $H_c$ and its corresponding $h_{inf}$ define the point of film coalescence.

However, we propose that, depending on the size of fluctuations, coalescence can be triggered at larger pore height, i.e., before reaching $H_c$ predicted by the original model. This occurs through a first-order–like film-thickening process, for example, analogous to liquid nucleation in binary mixtures of volatile and involatile components [42]. A detailed discussion on such process is beyond the scope of this Letter; here we only discuss a simple case where this transition becomes thermodynamically accessible. Consider the height at which the free energy difference between the metastable state $\Delta G_{21}(h_{min}, H)$ and unstable state $\Delta G_{21}(h_{max}, H)$ falls below $kT$ per unit area of the film as the characteristic thermal fluctuation energy [Fig. 3(a) and (c)]. That is to say, within a region where $\Delta G_{21}(h_{min}, H)$ differs from $\Delta G_{21}(h_{max}, H)$ by less than $kT$, films with different thickness values are energetically equivalent, analogous to the nucleus region described by Zeldovich [42]. Under this condition, the films on both sides abruptly thicken beyond $h_{max}$ and close their distance. Figure 3(d) shows that both models agree well with the experiment, though overall our model matches slightly better, especially at higher humidity. This fluctuation-driven onset of meniscus formation gives rise to a stochastic distribution of $H_c$, consistent with the spread observed experimentally in Fig. 2(a).

It is emphasized that even if coalescence is initiated as described by our model, the free energy barrier at $h_{max}$ pertains only to the film thickening and not to the formation of the meniscus. Once the distance between the surfaces falls below a critical value, the subsequent merging of the opposing films and meniscus formation becomes inevitable and lowers the system's free energy, and thus occurs spontaneously. In other words, the bridge itself is not a direct product of nucleation. This follows directly from $\Delta G_{31}$ for transition from (1) an empty to (3) a completely filled pore,

$$\Delta G_{31} = -\int_0^H \Pi(H')dH' - \frac{H}{v}kT\ln(S), \quad (8)$$

where $\Pi(H) = \Pi_0^H \exp(-H/H_m)$, for derivation see Appendix B. For sufficiently values of $H$ one finds $\Delta G_{31} < 0$ and $\Delta G_{31} < 2\Delta G_{21}$, for any admissible $h$. Here, admissible $h$ refers to values for which $H - 2h$ remains larger than molecular dimensions, thereby avoiding the divergence of the term $1/(H - 2h)^3$. By fitting $\Pi_0^H \exp(-H/H_m) = -\frac{kT}{v}\ln(S)$ to $H_c$ from our model [red curve in Fig. 3(d)]; $\Pi_0^H = 1299.8$ MPa and $H_m = 12.80$ Å are obtained (R-squared = 0.96) to approximate the effective disjoining pressure in the filled pore.

*The second scenario*—A substrate is dry when no stable film can exist under subsaturation, as on a hydrophobic surface where $\Pi(h) < 0$ for all $h$ [38,40]. When two such surfaces face each other in a slit pore, their attractive interactions reinforce each other, again resulting in a negative $\Pi(H)$ [43]. Consequently, no equilibrium film thickness $H_{eq}$ exists to fill the pore.

*Acknowledgment*—Ali Afzalifar is grateful to Prof. Dimo Kashchiev for helpful comments regarding the application of the nucleation theorem. Ali Afzalifar would like to extend thanks to Prof. Peter Liljeroth and Prof. Jaakko Timonen for advice on the experimental approach. This work is supported in part by the Research Council of Finland Center of Excellence Program (2022-2029) in Life-Inspired Hybrid Materials (LIBER) (project No. 346109), and Maj and Tor Nessling Foundation (decision No. 202200215), and the Research Council of Finland (grant No. 336557, and grant No. 323728).

*Appendix A: The case for applying the first nucleation theorem*—At fixed $S$, $T$, and $H$, the survival probability until time $t$ is

$$P(t) = \exp[-J(S)t], \qquad (A1)$$

where $J(S)$ (s$^{-1}$) is the time-independent nucleation rate at a given $S$ (and also $T$ which is dropped hereafter for brevity), yielding

$$J(S) = -\frac{\ln[P(t,S)]}{t}. \qquad (A2)$$

The first nucleation theorem depends only on $S$ as [31]

$$n^* \approx \frac{d\ln[J(S)]}{d\ln(S)} - \frac{d\ln[K(S)]}{d\ln(S)}, \qquad (A3)$$

where $K(S)$ (s$^{-1}$) is the kinetic prefactor, and an increasing function of $S$ meaning that $\frac{d\ln(K)}{d\ln(S)} > 0$. We therefore can define an upper-bound estimate

$$n^{*+} = \frac{d\ln[J(S)]}{d\ln(S)}, \qquad (A4)$$

in the sense that $n^*$ must be smaller than $n^{*+}$. Substituting $J(S)$ from Eq. (A2) into Eq. (A4) results

$$n^{*+} = \frac{d\ln[-\ln[P(t,S)]]}{d\ln(S)}, \qquad (A5)$$

However, in our measurements the tip moves with an average speed of $v_{tip} = 1.23$ (nm s$^{-1}$) towards the substrate to form the capillary bridge. In other words, while $S$ and $T$ are fixed, $H$ is not. Thus, the critical free energy $\Delta G^*(H,S)$ and $J(H,S)$ change with time. In fact, the system never stays long enough at any fixed $H$ to measure stationary nucleation rate. What is measured can be considered a cumulative (integrated) survival probability. For a time-dependent nucleation rate $J(t,S)$ we can rewrite Eq. (A1) as

$$P(t,S) = \exp\left[-\int_0^t J(t',S)dt'\right]. \qquad (A6)$$

Since $v_{tip}$ is constant, we use the tip height as a surrogate for time as

$$t' = (H_0 - H')/v_{tip}, \qquad (A7)$$

where $H_0$ is the tip height at $t = 0$. Using Eqs. (A6) and (A7), $P(t,S)$ can be represented as a $P(H,S)$

$$P(H,S) = \exp\left[-\frac{1}{v_{tip}}\int_H^{H_0} J(H',S)dH'\right], \qquad (A8)$$

where $P(H,S)$ is the cumulative survival probability from $H_0$ to $H$, which decays exponentially as $H$ decreases. Experimentally, we can measure $P(H,S)$ at different $S$ as

$$P(H,S) = \frac{M(H,S)}{M_{tot}}, \qquad (A9)$$

where $M(H,S)$ is the number of trials with no bridge formed until the tip reaches $H$ at a given $S$. From Eq. (A8)

$$-\ln[P(H,S)] = \frac{1}{v_{tip}}\int_H^{H_0} J(H',S)dH', \qquad (A10)$$

which allows calculating the derivative in Eq. (A5) as

$$\frac{d\ln(-\ln[P(H,S)])}{d\ln S} = \frac{d\ln\left[\frac{1}{v_{tip}}\int_H^{H_0} J(H',S)dH'\right]}{d\ln S} = \frac{S}{\frac{1}{v_{tip}}\int_H^{H_0} J(H',S)dH'} \cdot \frac{1}{v_{tip}}\int_H^{H_0}\frac{\partial J(H',S)}{\partial S}dH'. \qquad (A11)$$

From Eq. (A4) we have $\frac{S}{J}\frac{dJ}{dS} = n^{*+} \to \frac{dJ}{dS} = \frac{J n^{*+}}{S}$, therefore $\int_H^{H_0}\frac{\partial J(H',S)}{\partial S}dH' = \int_H^{H_0}\frac{n^{*+}J(H',S)}{S}dH'$. This allows rewriting Eq. (A11) as

$$\overline{n^{*+}} = \frac{d\ln[-\ln[P(H,S)]]}{d\ln(S)} = \frac{\int_H^{H_0} n^{*+} J(H',S)dH'}{\int_H^{H_0} J(H',S)dH'}, \qquad (A12)$$

showing that $\overline{n^{*+}}$ is a weighted average of $n^{*+}$ from $H$ to $H_0$, and it is experimentally accessible through $P(H,S)$. What Fig 2(c) shows is $\overline{n^{*+}}$ calculated by the first equality in Eq. (A12).

*Appendix B: Derivation of $\Delta G_{21}$ and $\Delta G_{31}$*—The Gibbs free energy of (1) a system containing $m$ vapor molecules in contact with two solid walls of a symmetric pore, with height $\ll$ width, is

$$G_v(m) = 2\sigma_{sv} + m\mu_v \equiv G_1, \qquad (B1)$$

where $\sigma_{sv}$ is the specific surface free energy of the bulk solid-vapor interface, and $\mu_v$ is the chemical potential of vapor molecules at temperature $T$ and pressure $p = Sp^s$, in which $p^s$ is the saturation pressure at $T$. Similar to the treatment for a single film laid down by Kashchiev [39], the Gibbs free energy of a pore system containing two thin films each having thickness of $h$ and $n$ molecules, and $m - 2n$ vapor molecules ($n \ll m$) in contact with the films is

$$G(m,n) = (m - 2n)\mu_v + 2n\mu_f^\infty + 2A\sigma(h) \equiv G_2, \qquad (B2)$$

where $\mu_f^\infty$ is the chemical potential of an infinitely-thick film, and $\sigma(h)$ is the specific surface free energy of the thin film. $\sigma(h)$ can be calculated knowing the disjoining pressure as [39]

$$\sigma(h) = \sigma_{sv} - \int_0^h \Pi(h',H)\, dh'. \qquad (B3)$$

Using Eqs. (B1)-(B3), the Gibbs free energy change per unit projected area of each film (i.e. for the half pore) in transition from (1) an empty pore to (2) a pore containing two adsorbed films $\Delta G_{21} = \frac{1/2}{A}(G_2 - G_1)$ is

$$\Delta G_{21} = -\int_0^h \Pi(h',H)\, dh' - \frac{n}{A}(\mu_v - \mu_f^\infty), \qquad (B4)$$

where by setting the saturation state as the reference state

$$\mu_v - \mu_f^\infty = kT\ln(S) - v(S-1)p^s. \qquad (B5)$$

Since $|v(S-1)p^s| \ll |kT\ln(S)|$, the final form of $\Delta G_{21}$ can be written as

$$\Delta G_{21} = -\int_0^h \Pi(h',H)\, dh' - \frac{h}{v}kT\ln(S), \qquad (B6)$$

where $h/v$ is equal to the number of molecules per unit area of each film, i.e. $h/v = n/A$.

The Gibbs free energy of (3) a completely filled pore with height $H$ is equivalent to that of a film with thickness of $H$ and $2n$ molecules filling the gap between two identical walls. For this film, similar to Eq. (B2), one can write

$$G(m,n) = (m - 2n)\mu_v + 2n\mu_f^\infty + A\sigma(H) \equiv G_3, \quad (B7)$$

where

$$\sigma(H) = 2\sigma_{sv} - \int_0^H \Pi(H') \, dH'. \quad (B8)$$

Therefore, $\Delta G_{31} = \frac{1}{A}(G_3 - G_1)$ is

$$\Delta G_{31} = -\int_0^H \Pi(H')dH' - \frac{H}{v}kT\ln(S). \quad (B9)$$